\documentstyle[epsfig,aps,prb,twocolumn]{revtex}
\begin{document}
\draft
\twocolumn[\hsize\textwidth\columnwidth\hsize\csname @twocolumnfalse\endcsname
\title{Friedel oscillations induced by  non--magnetic impurities\\
 in the two--dimensional Hubbard model} 
\author{ W.\ Ziegler, H. Endres and W.\ Hanke\\[4mm] }
\address{Institut f\"{u}r Theoretische Physik, Am Hubland,
D-97074 W\"{u}rzburg, Federal Republic of Germany\\[2mm]}
\maketitle

\begin{abstract}
%\noindent
We study the interplay of correlations and disorder using  
an unrestricted Slave--Boson  technique in real space.
Within the saddle--point approximation,
we find Friedel oscillations of the charge density
in the vicinity of a nonmagnetic impurity,
in agreement with numerical simulations.
The corresponding amplitudes are  suppressed by
repulsive interactions, while attractive correlations lead to a
charge--density--wave  enhancement.
In addition, we investigate the spatial dependence of
the local  magnetic  moment and the 
formation of a magnetic state at the impurity site. 
\end{abstract}
\vspace*{04mm}
]
\section{Introduction}
The combined effect of correlations and disorder, especially relevant
 in the context of  high--$T_c$ superconductors,
poses the theoretical
problem of  consistently  including both mechanisms
in a formal approach
\cite{Hirschfeld88,Hotta92,Borkowski94,Fehrenbacher94,Quinlan94,Hirschfeld94,Xiang94,scalapinorep}.
The combined influence of disorder and many--body interactions
can cause interesting new  physical processes,
and  leads to  unusual experimental properties.
In addition to the drastic
reduction of the transition temperature  in
cuprate superconductors, non--magnetic
impurities can, for example, create local magnetic moments
\cite{Xiao90,Mahajan94,Fink90,Alloul91,Walstedt93,Cieplak92,Ishida93}
which  extend  over up to several lattice spacings 
and  enrich the phase diagram
by introducing new disordered magnetic regions
\cite{Mendels94}. \\
Electron density (Friedel) oscillations as a response to a local
perturbation, i.\ e.\ the impurity, provides a clue to understanding
the electronic properties (Fermi surface etc.) of a many--electron system.
The period of the oscillation is determined by the Fermi surface, which may be defined by the
singular points in the momentum distribution function. Especially in connection with the
Fermi surface properties of high--temperature superconducting compounds \cite{Shen}
such as
$\rm Bi_{2}Sr_{2}CaCu_{2}O_{8+\delta}$, the Fermi surface evolution as a function of
doping from the under--doped to maximally and over--doped regimes is an extremely 
important, yet unresolved issue \cite{Ding}.
A systematic study of Friedel oscillations, both in experiment and in the generic
2--D Hubbard model for the high--$T_c$ cuprates, could, in principle, provide the
required information on the Fermi surface and how it changes from the extremely--low
doped case (with the possibility of "hole pockets", i.\ e.\ "small"
Fermi surfaces) to the larger doped situation with a large  Fermi surface,
with the latter being in accordance with Luttinger's theorem.\\
In a previous publication \cite{imp96}, a diagrammatic scheme  
was proposed that  uses the T--matrix formulation to
derive an  effective  scattering potential which includes the
effect of correlations as well as the bare scattering potential.
As the first step of a systematic approach, we focus here
on the one--particle effects of potential 
scattering in  correlated systems by formulating
a static  ansatz using an unrestricted Slave-Boson (USB) method,
which corresponds to the  effective--medium
renormalization of the scattering potential. \\
The calculations presented here  are done for the slightly overdoped 
case, i.e. $17\%$ doping away from half filling (i.\ e.\ the insulating
case for the 2--D Hubbard model). In this case, we observe the $2 k_F$--oscillations,
essentially unrenormalized from the ones of the non--interacting case. Not
only periodicity, but also the amplitude of the Friedel oscillations are shown here to be
in excellent agreement with Quantum Monte Carlo (QMC) simulations.
Corresponding calculations and comparisons with QMC for smaller doping 
are presently carried out, in order to test the simple effective medium 
approach also in the most interesting doping regime.

\section{Formalism}
We start with the Slave--Boson technique  as formulated
by Kotliar and Ruckenstein \cite{KR}.
An elementary introduction to the Slave--Boson formalism
for disorder--free
applications can be found, for example, in the review articles
by Arrigoni \em  et al.\em\cite{Arrigoni} and
Dieterich \em  et al.\em\cite{sb97}.
The canonical transformation 
\begin{equation}
c_{i\sigma}^{\dagger}\rightarrow
 (d^{\dagger}_i p_{i-\sigma}+p_{i\sigma}^{\dagger}e_i)
                       f_{i\sigma}^{\dagger}
\equiv
\tilde{z}_{i\sigma}^{\dagger}f_{i\sigma}^{\dagger},
\end{equation}
of the fermionic annihilation (creation) operator
$c_{i\sigma}^{(\dagger)}$  introduces new bosonic operators
$e_i^{(\dagger)},p^{(\dagger)}_{i\sigma},d^{(\dagger)}_i$
on a site $i$, which  can be
empty, singly occupied, or doubly occupied. Within the physical
subspace
%, in which the (pseudo--) fermionic operators  
%$f_{i\sigma}^{(\dagger)}$  are coupled to their  bosonic "slaves",   
the  Hamiltonian can be written as
\begin{equation}
   H_{SB}\,=\,
   \,-t
   \sum\limits_{\stackrel{\langle i,j \rangle}{\sigma}}
    \tilde{z}_{i\sigma}^{\dagger}f_{i\sigma}^{\dagger}
    f_{j\sigma}^{ }\tilde{z}_{j\sigma}^{ }
    +
    U\sum\limits_{i}d_i^{\dagger}d_i^{ }
    +
     V_0 \sum\limits_{\sigma}f_{0\sigma}^{\dagger}f_{0\sigma}^{ }
    .
    \label{HubbSB}
\end{equation}
Here the impurity is modeled  by a one--particle potential $V_0$
acting at the site $\vec{r}=0$ of  the lattice.
This model can be used, for example, to describe non--magnetic
scatterers in high--$T_c$ materials \cite{Fehrenbacher95}.\\
%
%
%
%
%An elementary introduction to the Slave--Boson formalism
%for disorder--free
%applications can be found, for example, in the review articles
%by Arrigoni \em  et al.\em\cite{Arrigoni} and 
%Dieterich \em  et al.\em\cite{sb97}. 
%
In the paramagnetic (spin--independent)
saddle-point approximation of the path--integral,
one calculates the partition function 
${\cal Z}_{Stat}=e^{-S_{Stat}}$  by
determing the stationary value of the effective action  
\begin{eqnarray}
S_{Stat}&=&-{\rm Tr}\,{\rm log}
       \left[\delta_{ij}(\partial_{\tau}-\mu+V_0\delta_{i0}+\lambda_i^{(2)})
   -t\,z_i
 z_j\right] \nonumber \\
   &&+\beta \sum_{i}
         \{d_i
(\lambda_i^{(1)}-2\lambda_i^{(2)}+U)d_i
         -\lambda_i^{(1)}
           \nonumber\\
        &&+  2p_i
(\lambda_i^{(1)}-\lambda_i^{(2)})p_{i} +
          e_i
\lambda_i^{(1)}e_i \}_{{}_{} },
\label{SeffUSB}
\end{eqnarray}
which is  given by the saddle--point equations, i.\ e.
\begin{equation}
\frac{\partial S_{Stat}}{\partial b_i}=0,
\quad b_i\in 
\{e_i,p_i,d_i,\lambda_i^{(1)},\lambda_i^{(2)} \}.
%\quad (i=1,...,N).
\label{seq}
\end{equation}
The derivatives must be calculated with respect to all
site--dependent bosonic variables and all Lagrange multipliers
$\lambda_i^{(1/2)}$, which are introduced in order to restrict
the enlarged, mixed bosonic--fermionic Hilbert space to the physical
subspace within the functional integral \cite{sb97}.
Note that we use  the standard renormalization factors 
$z_i =\frac{1}{\sqrt{1-d_i^2-p_i^2}}\tilde{z_i}\frac{1}{\sqrt{1-e_i^2-p_i^2}}$
 in Eq.\ (\ref{SeffUSB}) to guarantee the correct
$U\rightarrow0$ limit \cite{KR}.
On a $10\times 10$ lattice and
working at fixed average particle density $n$, one obtains
$85$ self--consistency equations
by exploiting the spatial symmetry of the bosonic variables which are
directly connected to  physical observables.  
%(\ref{seq}).
Within the Slave--Boson formalism, 
the particle density and the local magnetic moment are 
given by
\begin{equation}
n(\vec{r}_i)= 2d_i^2+2p_i^2,
\end{equation}
 and
\begin{equation}
m(\vec{r}_i)=2p_i^2.
\end{equation}
Compared to   mean--field  Slave--Boson approaches, which
can be restricted to 
homogeneous, spiral or bipartite configurations,
it takes  considerably
more  numerical effort to find the extreme value of $S_{Stat}$ in
the  high dimensional phase space encountered  here. 
The  strength of the Slave--Boson technique is
that many--particle  correlations are included  via
the $z$--factors in the effective masses of the quasi--free fermions.
For the  broken translational symmetry
considered here,
this renormalized  mean--field method \cite{Lilly90} 
also takes into account \em  site--dependent \em hopping amplitudes 
$t_{ij}^{eff}=t\,  z_i  z_j $.\\
By examining the fermionic propagator $G_f$ occurring in 
Eq.\ (\ref{SeffUSB}), one obtains the Dyson equation
\begin{eqnarray}
G_f^{-1}&=&-\left[(\partial_{\tau}-\mu)\delta_{ij}
    -t\, z_i^{\ast}z_j
  \right]
-(V_i\delta_{i0}+\lambda_i^{(2)})\delta_{ij}\nonumber\\
&=&G_{0f}^{-1}  - \Sigma^{USB},
\label{gpsfsig}
\end{eqnarray}
where the self energy is defined by 
\begin{equation}
\Sigma^{USB}= (V_i\delta_{i0}+\lambda_i^{(2)})\delta_{ij}.
\end{equation}
In an effective--medium or
Hartree--like  description of the scattering process
one finds \cite{imp96,pre} 
\begin{equation}
\Sigma^{Hartree}= (V_i\delta_{i0}+U\langle n_i\rangle)\delta_{ij},
\end{equation}
with site--dependent particle densities 
 $\langle n_i\rangle $.
Using the non--interacting propagator
$G_{0f}^{-1}=- (\partial_{\tau}-\mu)\delta_{ij} -t\, z_i^{\ast}z_j$,
Eq. (\ref{gpsfsig})
can be represented diagrammatically as in Fig.\ 1 
(compare also to Fig.\ 1 of Ref.\ 17).
Therefore, the correspondence  of the  Hartree-- and the USB--method
can be seen in the additional one--particle potential due to the
interplay of correlations and disorder on the  static
``mean--field'' level considered here.

\section{Results}
In the past, the Slave--Boson method has been
used as a powerful tool to calculate local quantities,
which have turned out to be in excellent agreement with numerically
exact results  \cite{Lilly90,Fresard92,sri96}. 
In order to check the reliability of the USB--method  in the
context of the extremely large system of coupled saddle--point
equations, 
we compared our results to
Quantum--Monte--Carlo  calculations. The QMC--simulations suffer
here from the problem 
of   increasing numerical instability 
due to the broken translational symmetry, which limits
the system size and the lowest accessible temperatures. 
Fig.\ 2  shows the comparison of the USB--data with QMC results for
an $8\times 8$--lattice. 
The charge density for $U=4t$ 
(in the following, we set $t=1$) is plotted as a function of the distance $r/a$
from the impurity site, where $a$ is the lattice constant.
  The saddle--point solution is almost identical
to the QMC result despite the fact that the latter
calculation is carried out at a
 relatively high temperature ($\beta=6, T=1/6$).
%
%At low temperatures, previous works have shown that
%Slave--Boson calculations give good agreement with
%numerical simulations \cite{Fresard92,preuss94}.\\
%
Fig.\ 3 displays  the USB oscillations of the
charge density in the vicinity of the impurity at an inverse temperature
$\beta=100$ for various values of the correlation strength U.
The solid line represents the non--interacting system and exhibits
the well--known $2 k_F$--oscillations, corresponding here to a
wavelength of $\lambda \approx 1.5a$.
While  the $2 k_F$--periodicity is unaffected\cite{anm},
repulsive interactions diminish the amplitude of the
 Friedel oscillations. They lead for strong correlations ($U>10$)
eventually to a short--range, monotonically decaying  behavior. 
The reason for this is that for increasing U the renormalized
(pseudo--fermionic) particles with  
correlation--enhanced effective masses
become more and more ineffective in contributing to
screening processes.
For attractive interactions, the impurity locally  breaks 
particle--hole symmetry and causes large Friedel
amplitudes.
This is due to the tendency of the attractive Hubbard 
model to form a  charge--density--wave
instability, which occurs in the
unperturbed system  at half--filling \cite{Lilly90_2,Moreo}. 
The oscillations present at $r/a>5$ are due to interference effects
at the corners of the $10\times 10$ lattice. They
are irrelevant for, for example, averaged random impurity sites. \\
Fig.\ 4 exhibits the influence of the potential
strength $V_0$ on the corresponding charge density behavior. Increasingly
attractive impurities enhance the amplitude until the maximum 
density ($n\rightarrow2$) is reached at the impurity site 
$r=0$ (inset of Fig.\ 4). Stronger potentials   decrease
the energy of the bound states below the quasi-particle band. 
A  symmetric  repulsive--attractive effect  of the impurity 
can  be found in the  oscillations for $V_0=\pm1$
around the average density ($n=0.81$).
Further USB--calculations showed that higher temperatures 
cause a decrease of Friedel oscillations, which is in accordance
with  the smearing out
of the jump in the Fermi function at $k_F$,  which determines, for
example, the  integral bond in a continuum calculation  \cite{Mahan}
of $n(r)$.
This is also the origin of the moderate oscillations in Fig. 2
at $\beta=6$.
We note that no significant deviations from the
above described behavior were found in the investigated metallic phase
at $0.7 < n < 0.9$.\\
We now  concentrate on the one--particle spectra obtained from the
USB--calculations.
Using the pseudo--fermionic propagator,  we obtain the density of states
\begin{eqnarray}
D(\omega)&=&-\frac{1}{N\pi} \sum_m {\rm Im}\, G_f(m,\omega+i\eta),
\nonumber \\
& =& \frac{1}{N}\sum_m \delta(\omega-\varepsilon_m).
\end{eqnarray}
This function, with the $\delta$--functions replaced by  $\eta$--broadened
functions, is plotted in Fig.\ 5. 
The broken translational symmetry lifts the  degeneracy of various 
quasi-particle states of the unperturbed system (solid line) and
produces a single bound state 
(arrow in Fig.\ 5) separated from the band.   
Note that this band has a smaller width than in the non--interacting
case due to the Slave--Boson  renormalization effects.
The bound state possesses
$s$--wave symmetry (see, for example, the first state in Fig.\ 6),
as does   the bare scattering potential $V_0$.
Therefore, within our  static potential renormalization
\cite{pre}, the 
effective  impurity potential \cite{imp96}
is not  strong enough to 
cause  the previously proposed 
(and verified in numerical simulations \cite{Imptj})
anisotropic scattering processes, resulting,
for example in bound states of higher symmetries. 
These extended bound states can be found by
 introducing  an  additional bare one--particle potential $V_1$
at the neighboring sites, which
produces states of all the possible  irreducible representations \cite{imp96}.
\mbox{Fig.\ 6} shows the corresponding   amplitudes of  the pseudo--fermionic
(bound state)  wave-functions
$ |\Psi_m\rangle=\sum_{i}c_{mi}f^{\dagger}_{i}|vac\rangle$, 
belonging to the six lowest lying energies $\varepsilon_m$. \\
Another interesting feature, found in the USB--results, occurs
in the magnetic structure influenced by the non--magnetic impurity.
We first investigate the local magnetic moment $m(0)$ at the impurity
site $\vec{r}=0$. In Fig.\ 7, this quantity is plotted  (for fixed 
$U$ and average particle density $n$)  as a function of the
potential strength $V_0$ and  the corresponding local particle density
$n(0)$.
The maximum of  $m(0)$ at $V_0\approx -1.0$
corresponds to a locally  half--filled impurity
site, i. e. $n(0)=1$,  and signals the formation of a magnetic state.
This situation is analogous to the mean--field description of
the Single--Impurity Anderson model 
and was also found in exact diagonalization studies
of a Hubbard chain containing a single impurity
by Hallberg and Balseiro \cite{Hallberg}.
In order to evaluate the  size of the magnetic moment,
one should perform spin--dependent USB--calculations, which could then
also be extended  to the  symmetry--broken phases found near half--filling
\cite{Lilly90,sri96}. 
The  paramagnetic subspace to which we restrict the solution appears 
 sufficient for the questions investigated here.
Substituting  $U+U_0$ for the local interaction strength $U$ at $\vec{r}=0$,
one  shifts the magnetic transition point by a
small $U_0$,  which can be seen in the inset of  Fig.\ 7. 
This means that
the appearance of a magnetic state depends on a fine tuning  of the system 
parameters, which could be one reason for the experimental disagreement
concerning the size and existence of a magnetic moment
\cite{Xiao90,Mahajan94,Fink90,Alloul91,Walstedt93,Cieplak92,Ishida93}.
Strong potentials eventually lead to the formation of a  $S=0$
spin--singlet (in the limit of no or double occupancy), which 
corresponds in Fig.\ 7 to a  minimal
local moment at the impurity site.
Interestingly,
the oscillations of the charge density are also  accompanied by
spatial variations of the magnetic moment. 
Using  the parameters relevant for Fig.\ 7, Fig.\ 8 shows the magnetic
structure in the vicinity of the impurity.
For $V_0=-1$, the magnetic state at $\vec{r}=0$ produces a relatively
small amplitudes  $m(r)$. The stronger attractive potential,
$V_0=-4,$  creates a singlet bound state at $\vec{r}=0$ and leads
to a maximal moment at a distance of $r=\sqrt{2}$. This 
can be connected to the experimentally detected spatially extended 
(para--)magnetic moments around $Zn$--impurities. 
In a theoretical picture, this  feature of the metallic
phase can be seen as 
a reminiscence of correlation--induced bound states, carrying localized
magnetic moments, at half--filling \cite{Imptj,ImptjII}.

\section{Conclusion}
In summary, we have extended the  Slave--Boson mean--field method
to systems with a spatially  inhomogeneous  phase space.
This technique can be used to calculate single--particle properties of 
 correlated--lattice models which include a non--magnetic impurity.
Formally, this method corresponds to a static renormalization of
the impurity potential in analogy to a diagrammatic 
effective--medium Hartree
approach to impurity scattering \cite{imp96}.\\
We have solved numerically  the  complicated
system of 85 coupled saddle--point equations for a $10\times10$
Hubbard model in the metallic phase.
The USB--results show, in accordance with QMC--data, Friedel
oscillations of the charge density in the vicinity of the impurity.
The amplitude is reduced by repulsive correlations, while the
attractive  system display a charge--density--wave enhancement.
The system parameters can be chosen,  so that a magnetic state
appears at the impurity site, signaled by the  maximal
local magnetic moment. In addition, the range of the magnetic
correlations around the impurity increases with increasing potential
strength, being in qualitative accordance with
experimental results.

The above  formalism and its results
can be viewed as  the first step of a systematic 
investigation of correlation effects on impurity scattering.
Higher diagrammatic orders of many--body interactions
will be studied  in a future publication \cite{pre}.\\

We would like to thank 
D. Poilblanc and D. J. Scalapino for  stimulating discussions
and 
P. Dieterich for technical help in  solving the  USB equations.
We also thank   the Bavarian
"FORSUPRA" program for  financially supporting this work.

\newpage 

\indent
%.............................................

%\newpage
\parindent0em
{\bf\large Figure captions}\\

FIG.\ 1:
Diagrammatic representation of the self--consistent
pseudo--fermionic one--particle
propagator $G_f$ (double line). 
The scattering by the impurity is depicted
by the cross, the   zig--zag
 line represents the Slave--Boson self--energy,
 i. e. the  Lagrange parameter $\lambda^{(2)}$.\\

FIG.\ 2:
Radial modification of the charge density  caused by a $\delta$--potential
$V_0=-8.0$  The data points of the Slave-Boson calculation fit the 
QMC--distribution for the Hubbard--model at $U=4$, $\beta=6$ and an 
average density  $n=0.83$.\\

FIG.\ 3:
Correlations effects on the low--temperature ($\beta=100$)
Friedel oscillations induced by a 
$\delta$--potential of strength $V_0=-4.0$.\\

FIG.\ 4:
USB--calculations for different impurity potentials in the Hubbard
model ($U=4$, $\beta=100$).  Repulsive (attractive) interactions
reduce (enhance) the amplitude of Friedel oscillations. \\

FIG.\ 5:
USB--spectra of the pseudo--fermionic propagator. The density of states
shows an impurity--induced $s$--wave bound--state below the 
renormalized quasi-particle band. The broken translational symmetry creates
states  between the unperturbed ($V_0=0$) single--particle energy levels.\\

FIG.\ 6:
Wave functions of the six lowest pseudo--fermionic eigenstates caused
by an extended impurity ($V_0=-8.0$, $V_1=-4.8$). The 
wave functions exhibit $s$--, $p_{x/y}$--
 (degenerated), $s$--, $d_{x^2-y^2}$-- and
$s$--wave symmetry, in order from upper left to lower right. \\

FIG.\ 7:
Charge density and magnetic moment   at the impurity site $(r=0)$. 
The scattering potential acts in the $U=4$  Hubbard model at
$\beta=100$. The maximum of $m(0)$  and the  half--filled
site displays the formation of a magnetic state.\\
 
FIG.\ 8:
The local magnetic moment  in the vicinity
of the impurity. The spatial  magnetic distribution and its maximum
depends on the strength   of the scattering potential $V_0$.\\

 \newpage
\begin{onecolumn}

FIG.\ 1\\
\begin{figure}[t]
\begin{center}
\epsfig{file=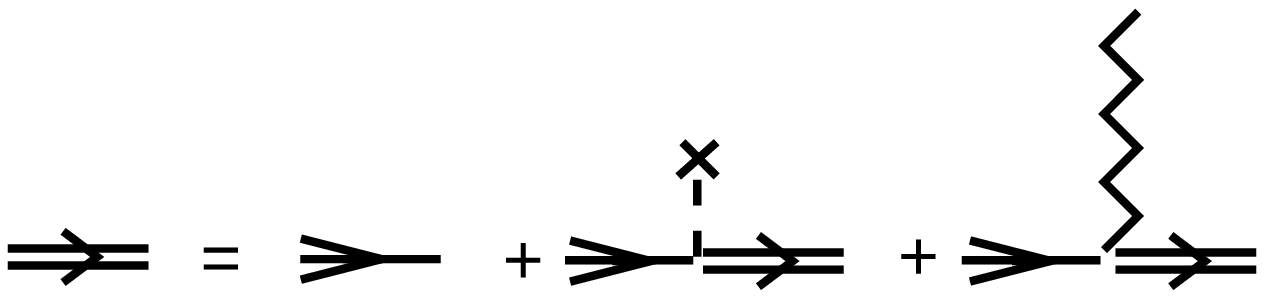,scale=0.5,angle=-90}\\[1cm]
\end{center}
\end{figure}

FIG.\ 2\\
\begin{figure}[h]
\begin{center}
\epsfig{file=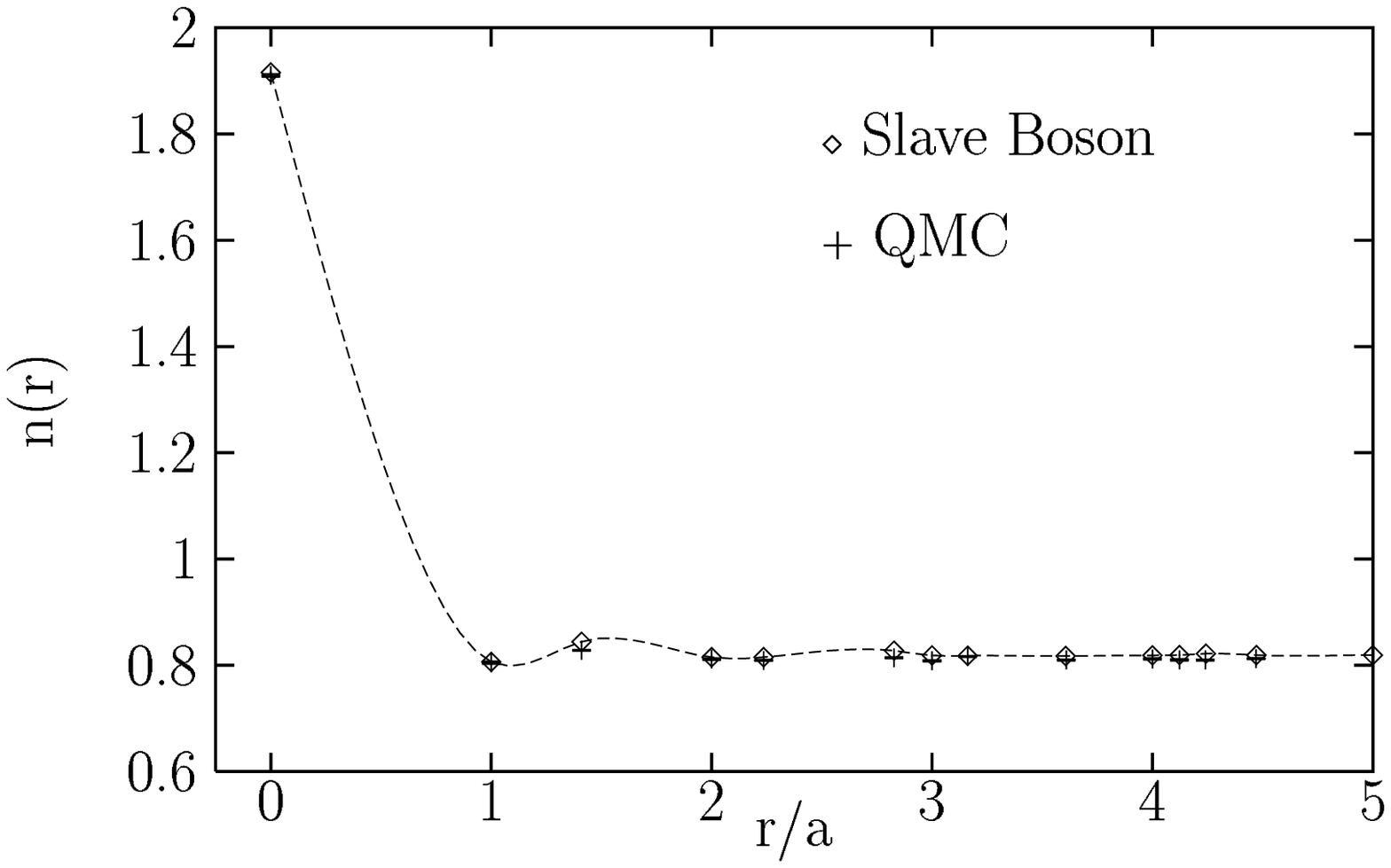,scale=0.9}
\end{center}
\end{figure}

\newpage

FIG.\ 3\\
\begin{figure}[t]
\begin{center}
\epsfig{file=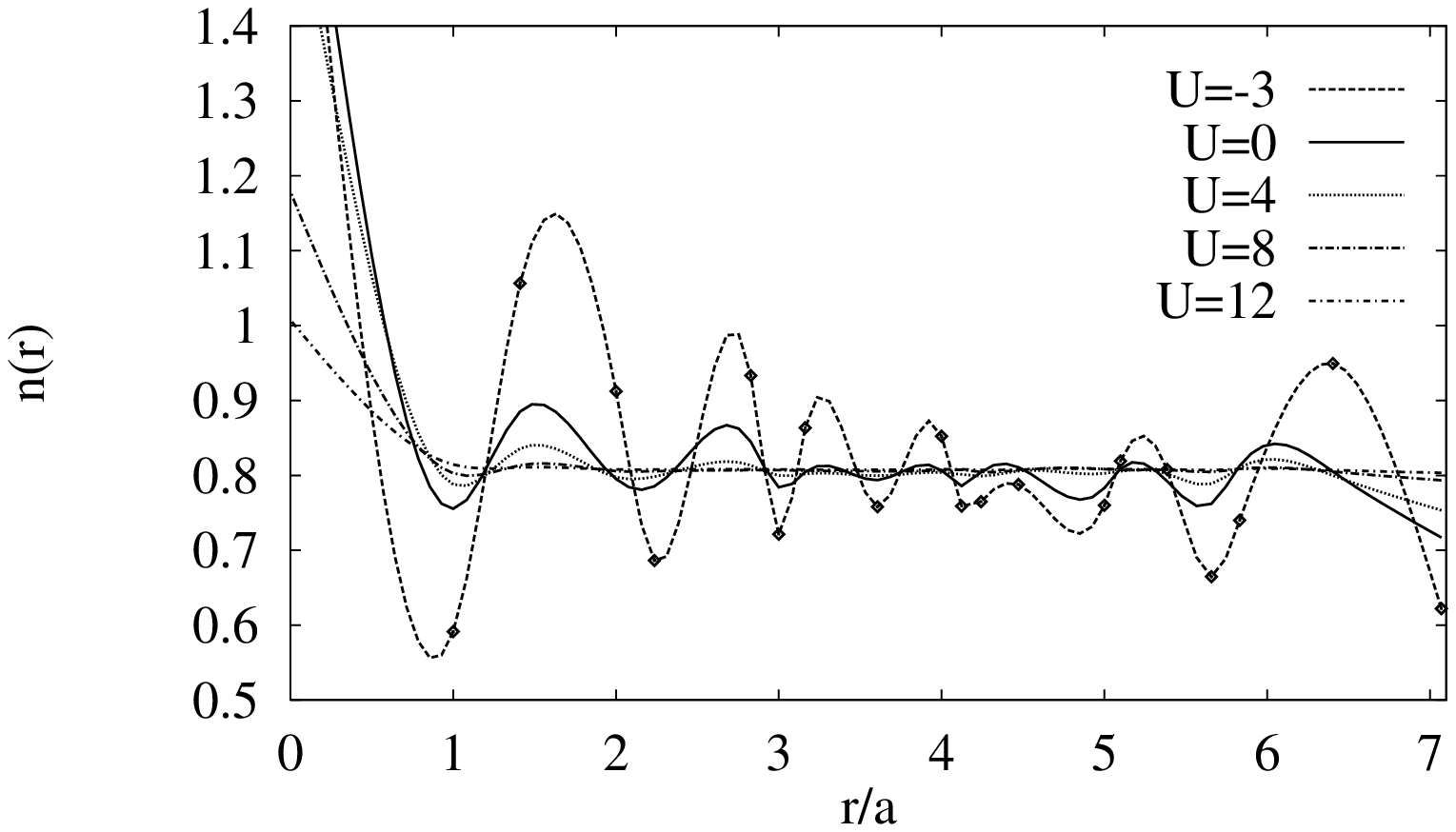,scale=1.0}
\end{center}
\end{figure}

FIG.\ 4\\
\begin{figure}[b]
\begin{center}
\hspace*{-0cm}
\epsfig{file=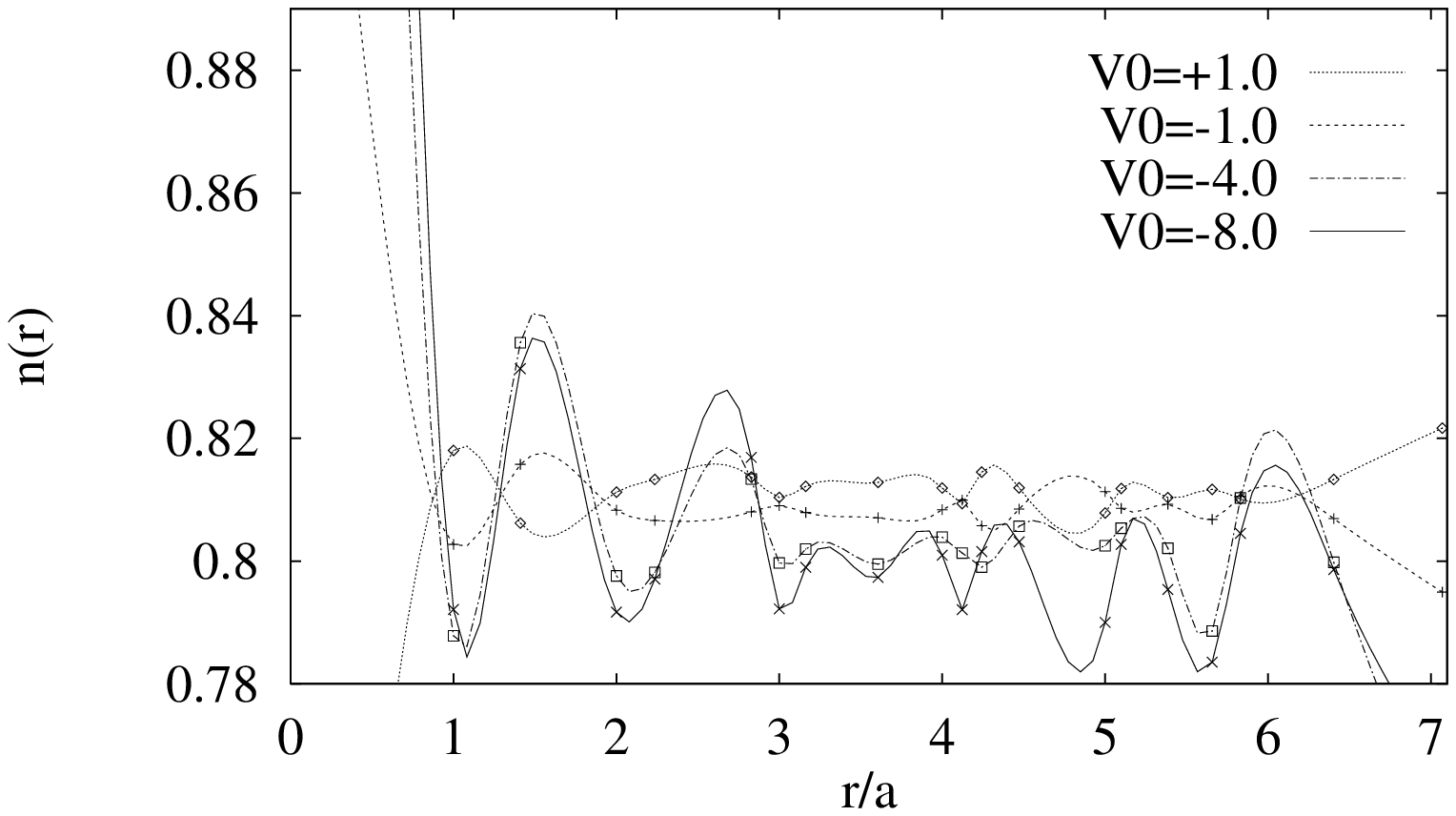,scale=1.0}

\vspace*{-8.5cm}
\hspace*{+7.1cm}
\epsfig{file=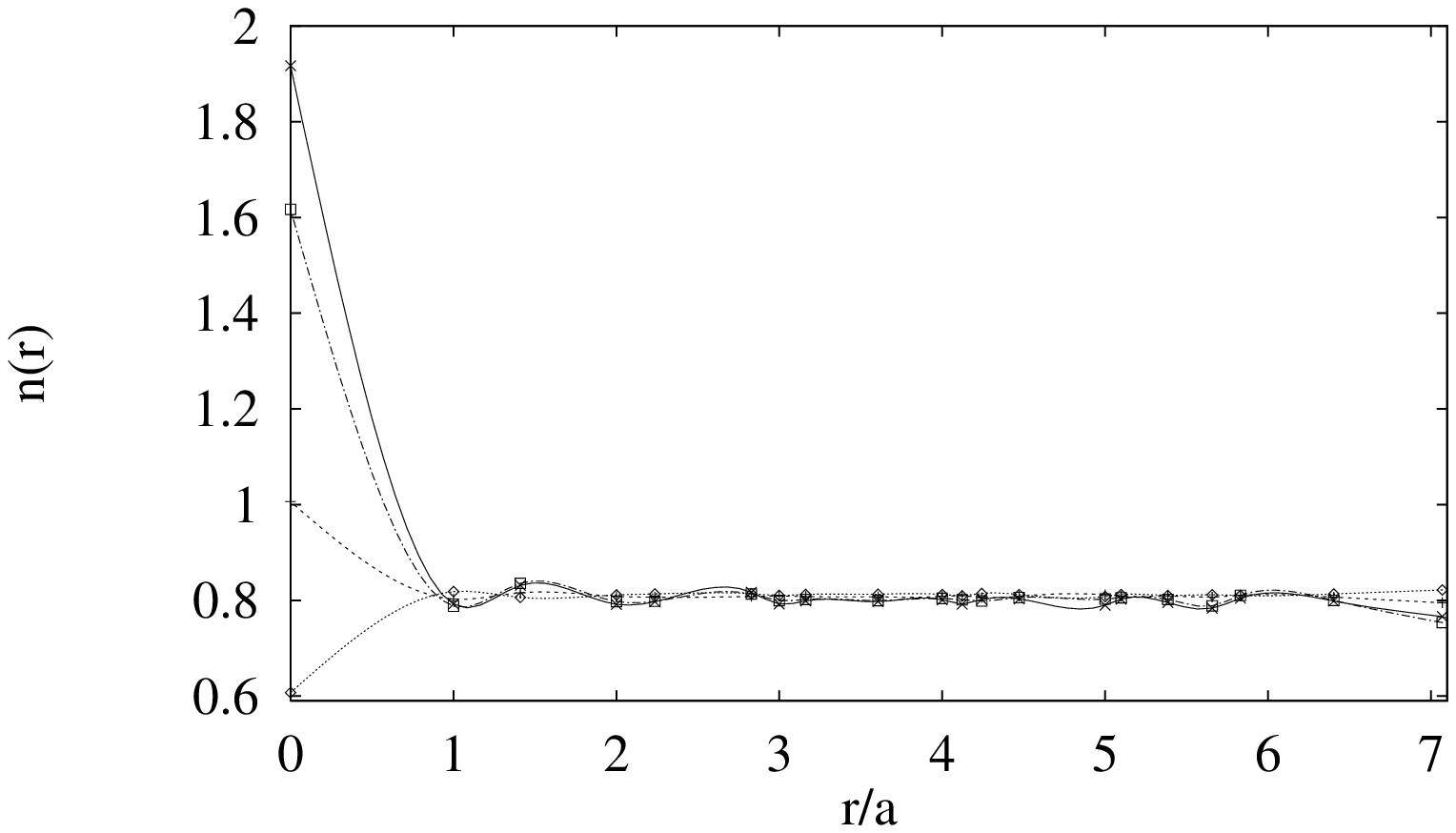,scale=0.5}
\vspace*{3cm}
\end{center}
\end{figure}

\newpage

FIG.\ 5\\
\begin{figure}[t]
\begin{center}
\hspace*{-1.0cm}
\epsfig{file=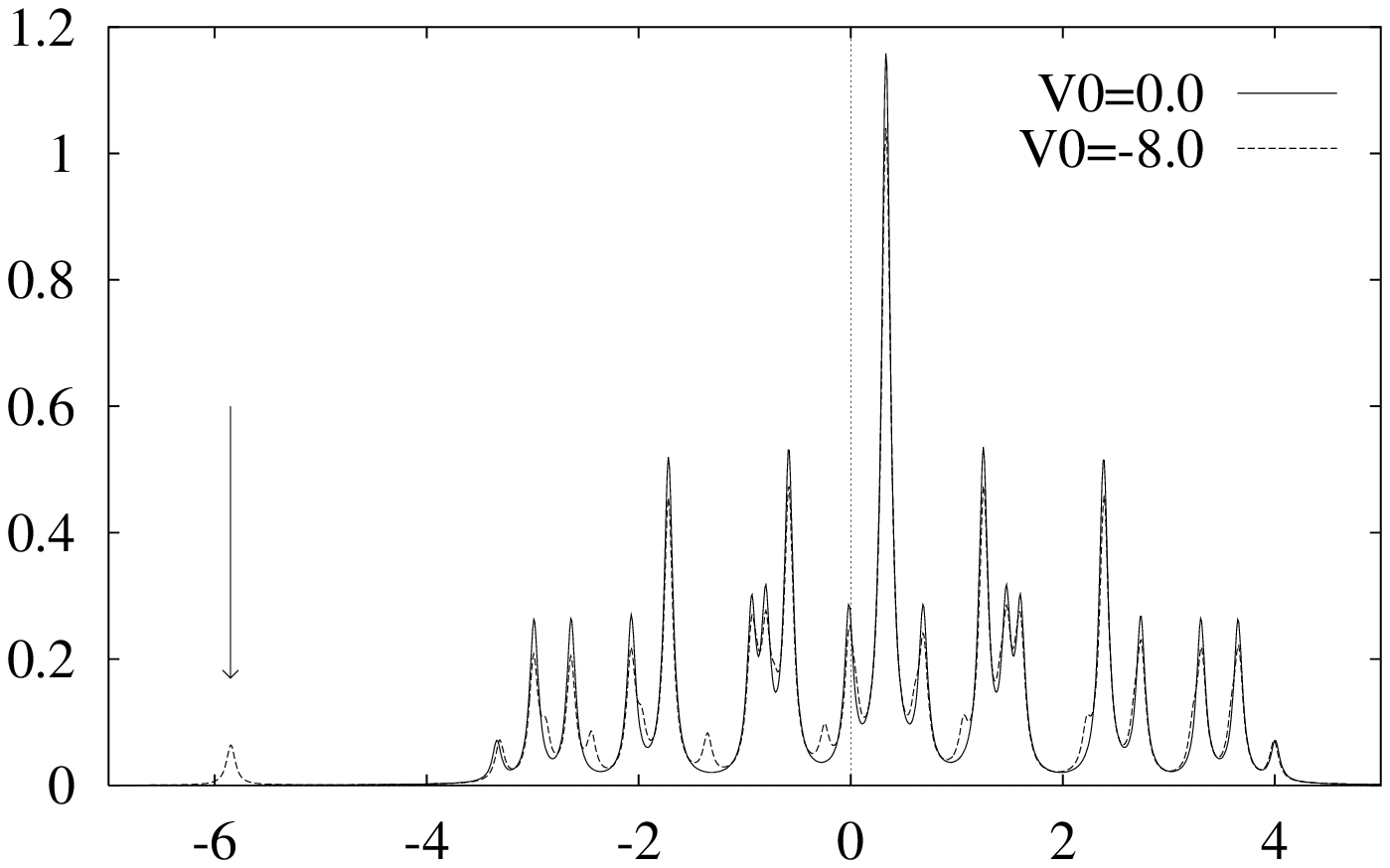,scale=0.8}
\end{center}
\end{figure}

%\newpage

FIG.\ 6\\
\begin{figure}[b]
\hspace*{-6mm}
\epsfig{file=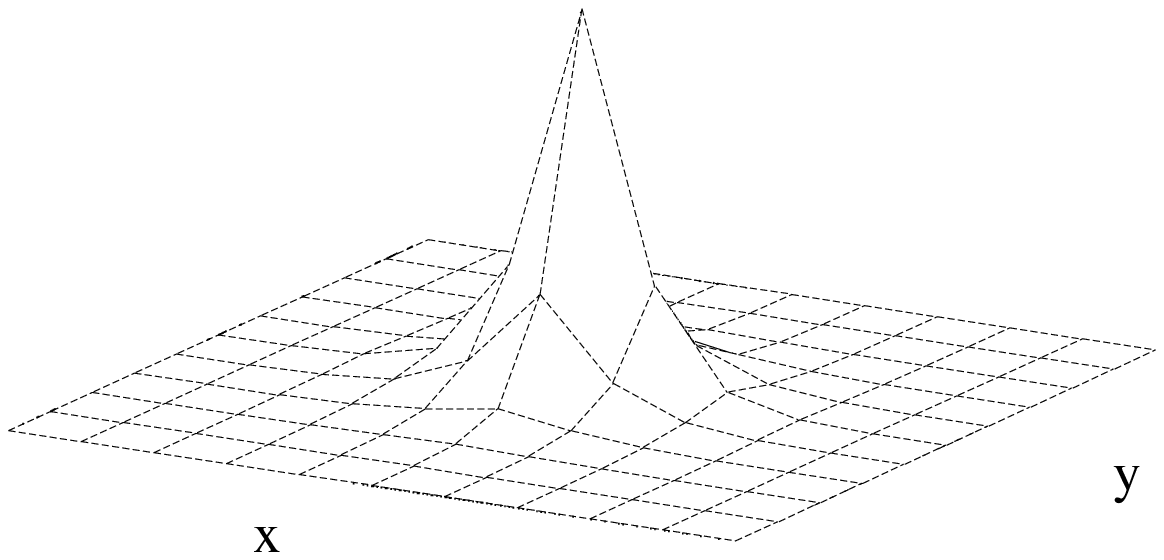,scale=0.4}
\hspace*{-1.6cm}
%\vspace*{-16mm}
\epsfig{file=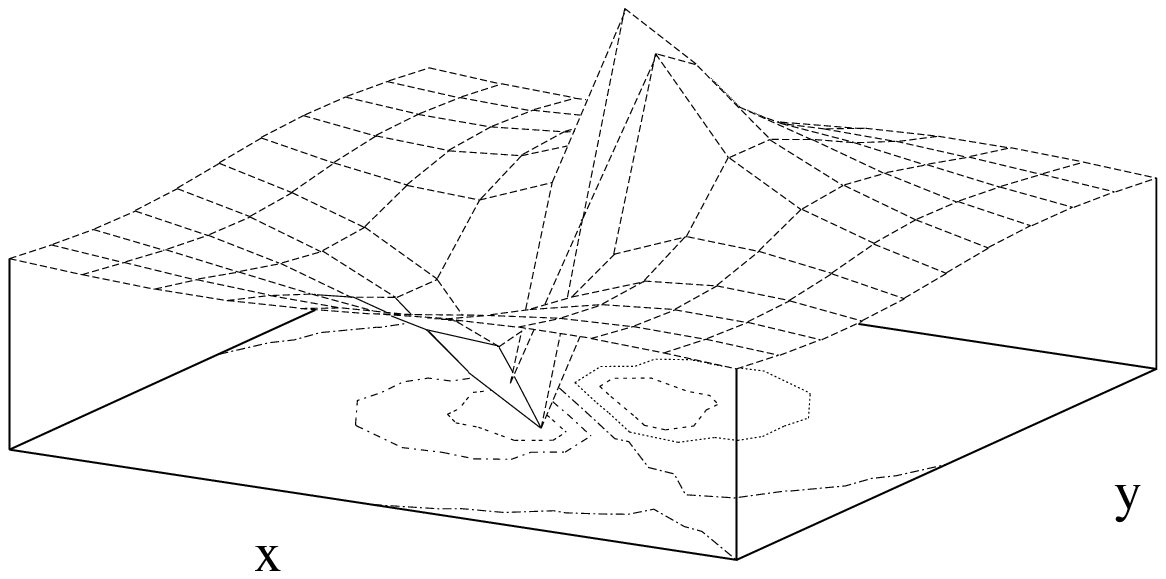,scale=0.4}
\hspace*{-9mm}
\epsfig{file=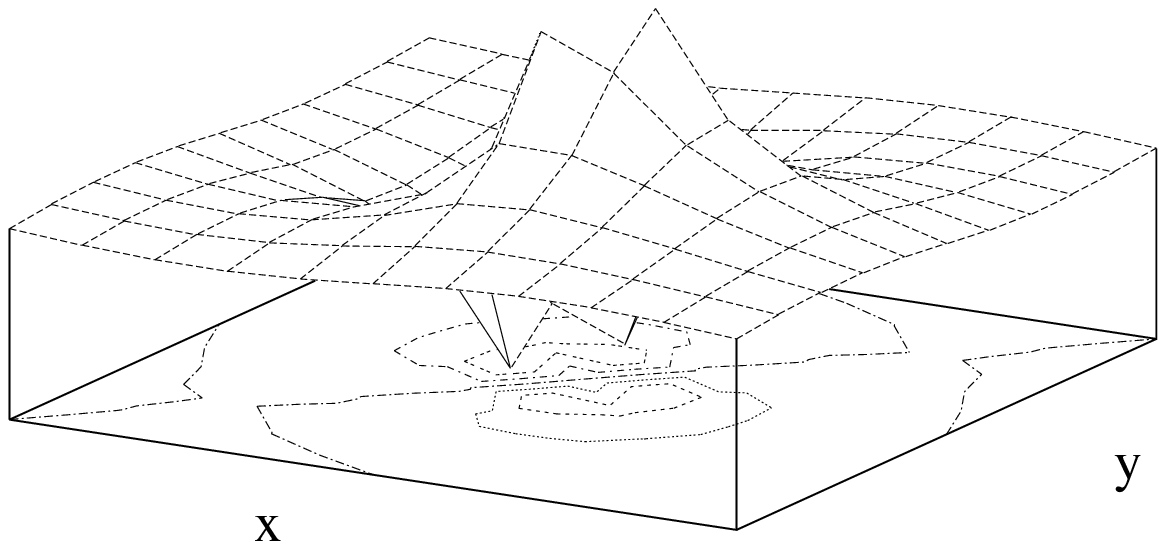,scale=0.4}
\newline
\hspace*{-7mm}
\epsfig{file=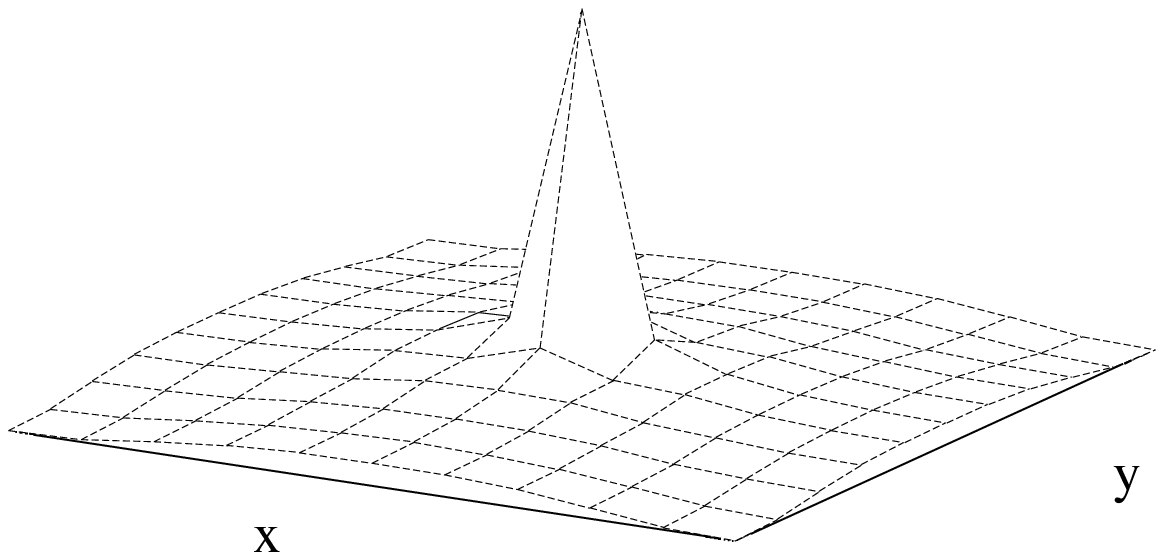,scale=0.4}
\hspace*{-16mm}
%\vspace*{-46mm}
\epsfig{file=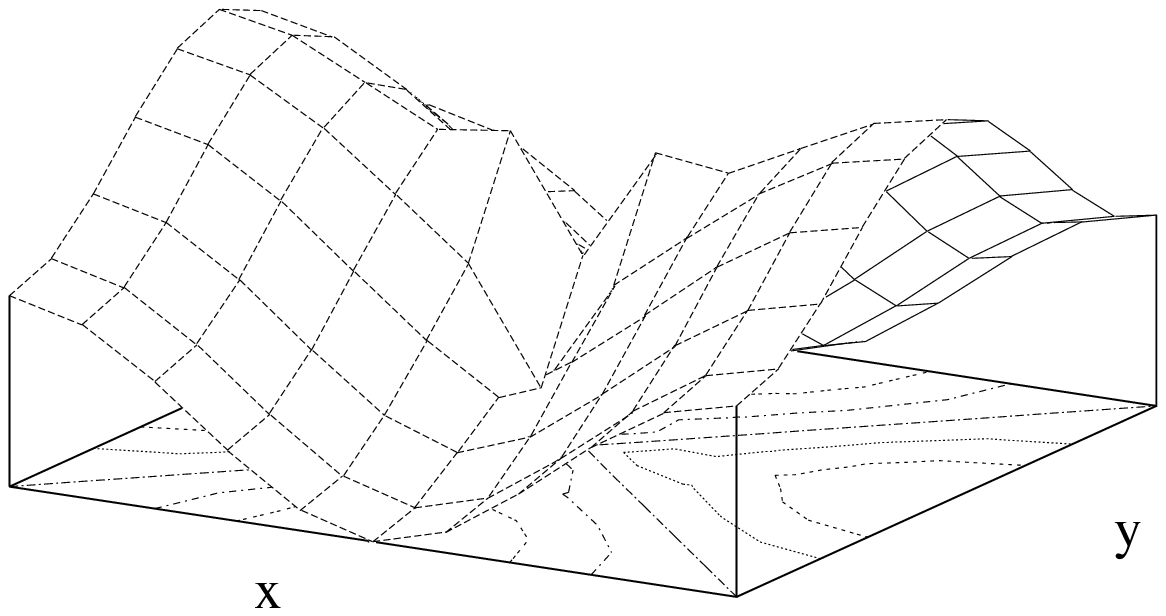,scale=0.4}
\hspace*{-9mm}
\epsfig{file=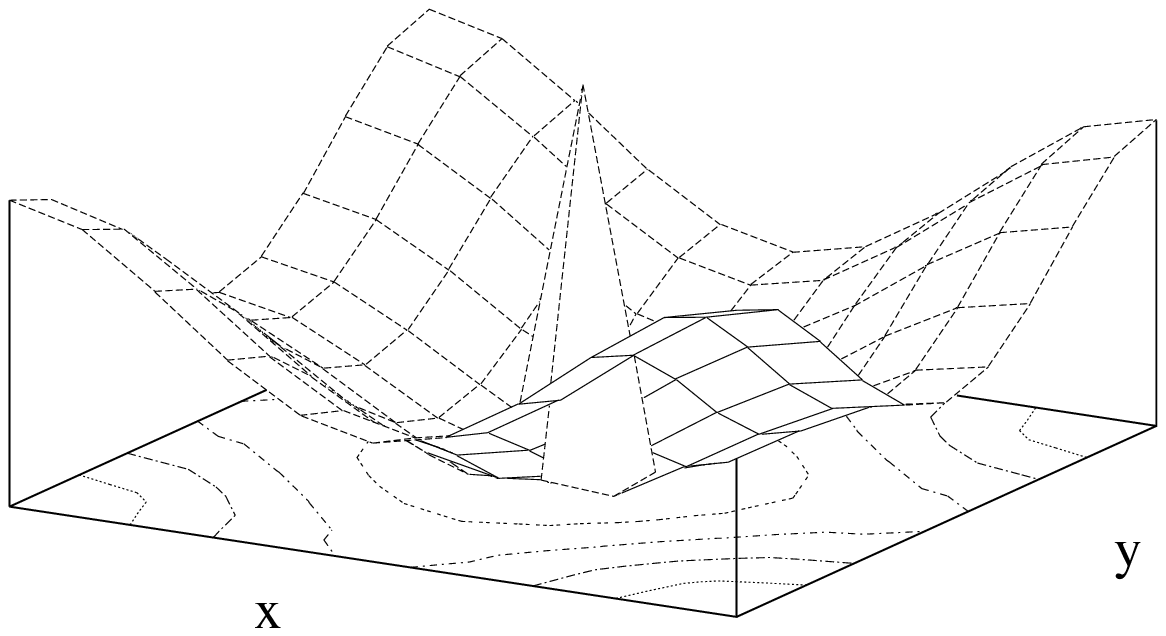,scale=0.4}
\end{figure}

\newpage
FIG.\ 7\\
\begin{figure}[t]
\hspace*{-1cm}
\epsfig{file=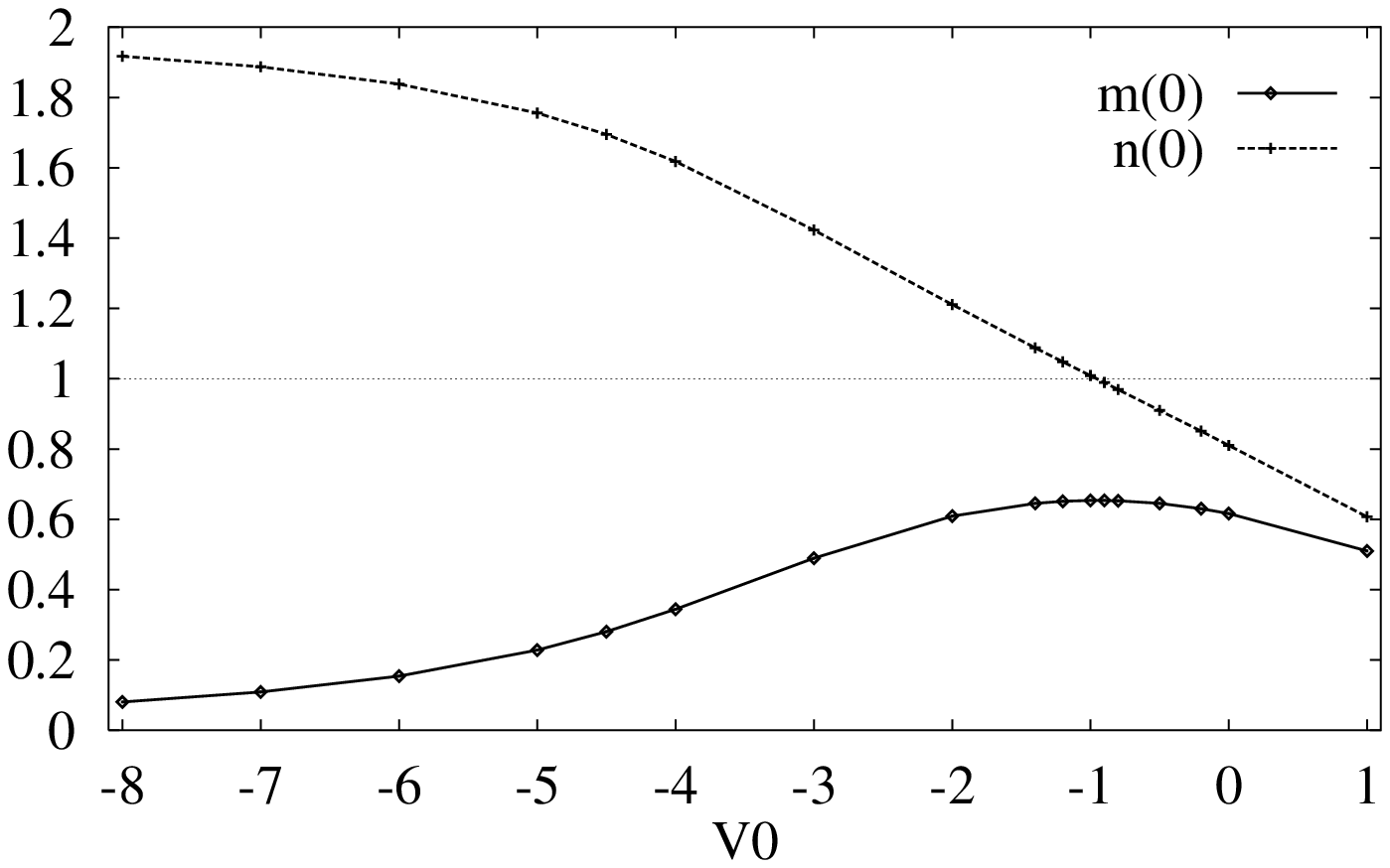,scale=1.0}

\vspace*{-6.72cm}
\hspace*{12mm}
\epsfig{file=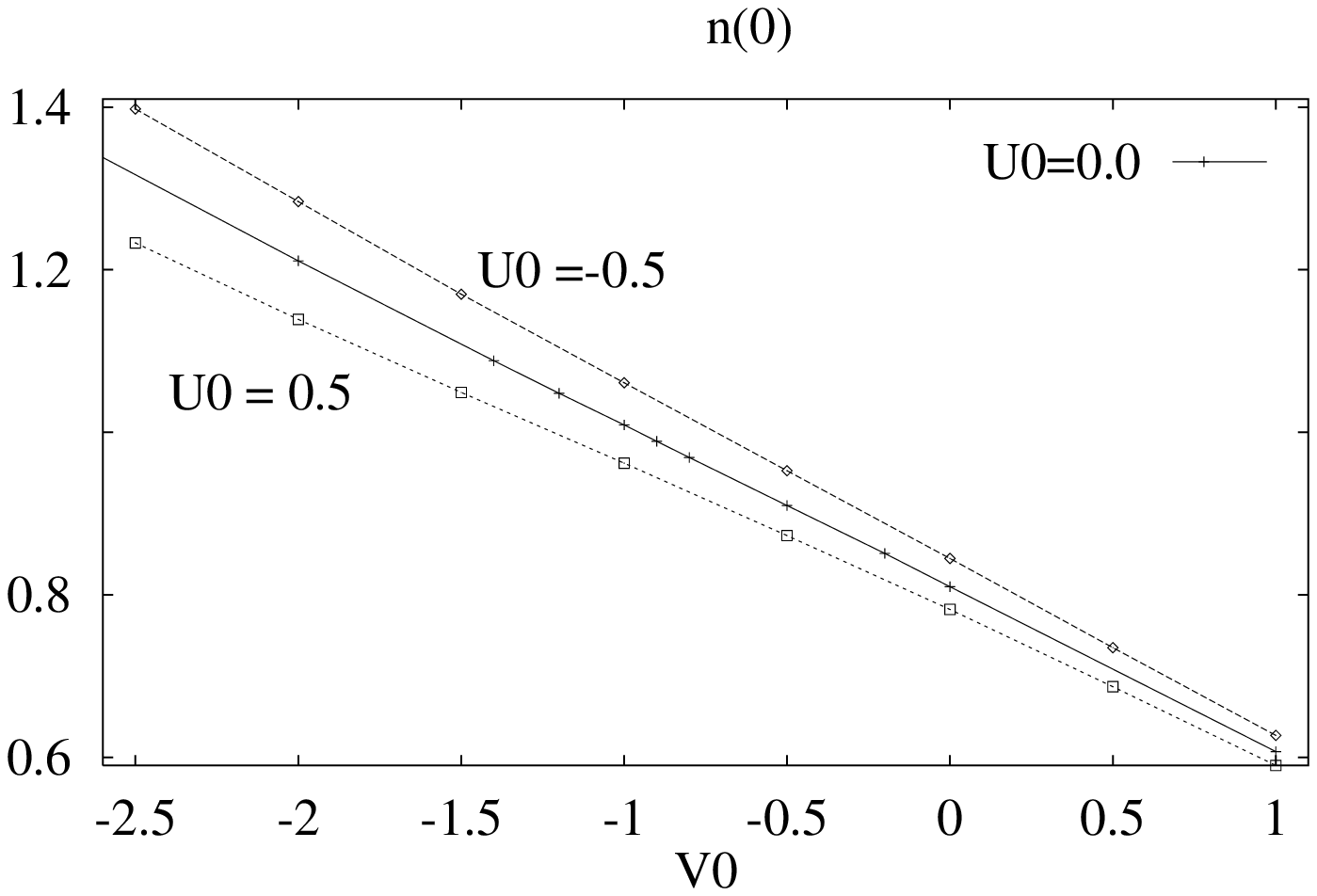,scale=0.441}
\vspace*{20mm}
\end{figure}

FIG.\ 8\\
\begin{figure}[b]
\begin{center}
\hspace*{-35mm}
\epsfig{file=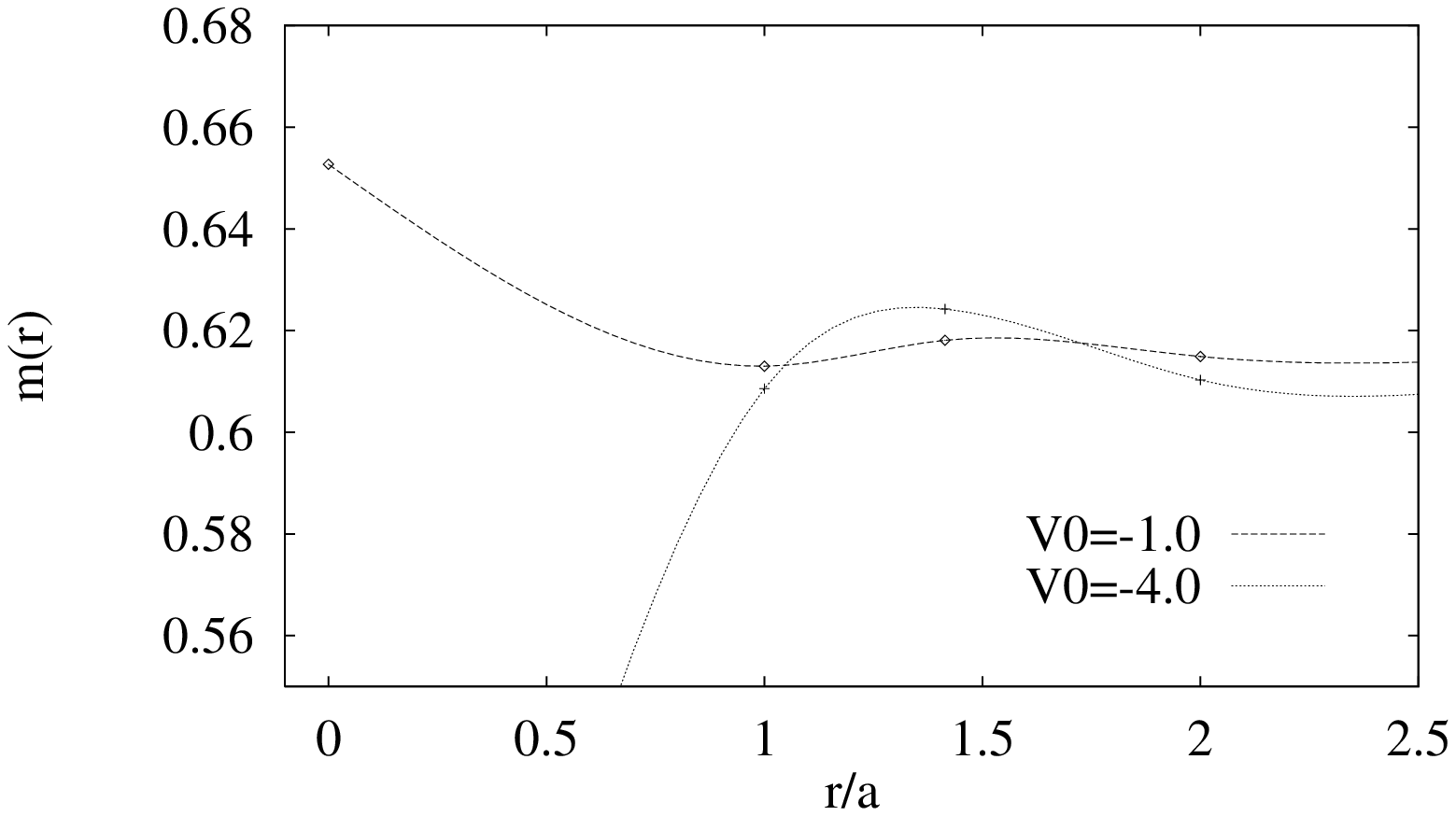,scale=0.99}
%\vspace*{-10mm}
\end{center}
\end{figure}

\end{onecolumn}

\newpage

\begin{thebibliography}{99}

\bibitem{Hirschfeld88}P.\ J.\ Hirschfeld, P. W\"olfle and D. Einzel, 
                   Phys.\ Rev.\ B {\bf 37}, 83 (1988).
\bibitem{Hotta92}T.\ Hotta, J.\ Phys. Soc.\ Jpn.\  {\bf 62},
                  274 (1993).
\bibitem{Borkowski94}L.\ S.\ Borkowski and P.\ J.\ Hirschfeld,
                   Phys.\ Rev.\ B {\bf 49}, 15404 (1994).
\bibitem{Fehrenbacher94}R.\ Fehrenbacher and M.\ R.\ Norman,
                    Phys.\ Rev.\ B {\bf 50}, 3445 (1994).
\bibitem{Quinlan94}S.\ M.\ Quinlan and D.\ J.\ Scalapino,
                   Phys.\ Rev.\ B {\bf 51}, 497 (1994).
\bibitem{Hirschfeld94}P.\ J.\ Hirschfeld, W.\ O.\ Puttika and
                    D.\ J.\ Scalapino,  Phys.\ Rev.\ B {\bf 50},
                    10250 (1994).
\bibitem{Xiang94}T.\ Xiang and J.\ M.\ Wheatley,  Phys.\ Rev.\ B {\bf 51},
                 11721 (1995).
\bibitem{scalapinorep} D. J. Scalapino, Phys. Rep.\ {\bf 250},
                        329 (1995).
%....................
\bibitem{Xiao90}G. Xiao \em et al.\ \em,
                 Phys.\ Rev.\ B {\bf 42}, 8752 (1990).
\bibitem{Mahajan94}A.\ V.\ Mahajan \em et al.\ \em,
                Phys.\ Rev.\ Lett.\ {\bf 72}, 3100
               (1994).
\bibitem{Fink90} A.\ M.\ Finkel'stein \em et al.\ \em,
                 Physica C {\bf 168}, 370 (1990).
\bibitem{Alloul91}H.\ Alloul \em et al.\ \em,
                  Phys.\ Rev.\ Lett.\  {\bf 67},
                  3140 (1991).
\bibitem{Walstedt93}R.\ E.\ Walstedt \em et al.\ \em,
               Phys.\ Rev.\ B, {\bf 48}, 10646 (1993).
\bibitem{Cieplak92}M.\ Z.\ Cieplak \em et al.\ \em,
                Phys.\ Rev.\ B {\bf 46}, 5536 (1992).
\bibitem{Ishida93}K.\ Ishida \em et al.\ \em,
                J.\ Phys.\ Soc.\ Jpn.\
                {\bf 62}(8), 2803 (1993).
\bibitem{Mendels94} P. Mendels \em et al.\ \em,
                Phys.\ Rev.\ B {\bf 49}, 10035 (1994).
\bibitem{Shen} Z.--X. Shen and D. S. Dessau, Phys. Rep. {\bf 253}, 1 (1995).
\bibitem{Ding} H. Ding \em et al. \em, Phys.\ Rev.\ Lett.\ {76}, 533 (1996).
%\bibitem{Chien91}T.\ R.\ Chien \em et al.\ \em,
%                Phys.\ Rev.\ Lett.\ {\bf 67}, 2088 (1991).
%\bibitem{Zhao93}Y.\ Zhao \em et al.\ \em,
%               J.\ Phys.\ Condens. Matter 5 (22), 3623 (1993).
%............................................
\bibitem{imp96} W. Ziegler, D. Poilblanc, R. Preuss, W. Hanke and
               D. J. Scalapino, Phys.\ Rev.\ B {\bf 53}, 8794 (1996). 
\bibitem{KR} G. Kotliar and A.\ E.\ Ruckenstein, Phys.\ Rev.\ Lett.\
             {\bf 57}, 1362 (1986).
\bibitem{Arrigoni}E. Arrigoni, C. Castellani, M. Grilli,
          R. Raimondi and G.\  C. Strinati, Phys. Rep.\ {\bf 241},
           291 (1994)
\bibitem{sb97} P. Dieterich, W. Ziegler, A. Muramatsu and W. Hanke,
               to appear in Phys. Rep. (1997).
\bibitem{Fehrenbacher95}  R.\ Fehrenbacher, Phys.\ Rev.\ Lett.
       {\bf 77}, 1849 (1996).
\bibitem{Lilly90} L. Lilly, A. Muramatsu and W. Hanke,
                  Phys.\ Rev.\ Lett. {\bf 65}, 1379 (1990).
\bibitem{pre}W.\ Ziegler, G. Hildebrand and W.\ Hanke, to be published.
\bibitem{Fresard92} R. Fr\'esard and P. W\"olfle, J.\ Phys. Cond. Mat.
                {\bf 4}, 3625 (1992).
\bibitem{sri96} W. Ziegler,  P. Dieterich, A. Muramatsu and W. Hanke,
                Phys.\ Rev.\ B {\bf 53}, 1231 (1996).
\bibitem{preuss94} R. Preuss, A. Muramatsu, W. von der Linden, 
                   P. Dieterich, F. F. Assaad and W. Hanke,
                    Phys.\ Rev.\ Lett. {\bf 73}, 732 (1994).
\bibitem{anm}  The anisotropic Fermi--surface of the
               lattice system causes the small deviations from 
               a harmonic oscillation in Fig.\  3.
\bibitem{Lilly90_2}L.\ Lilly, A. Muramatsu and W. Hanke,
                Phys. B {\bf 165\& 166}, 393 (1990).
\bibitem{Moreo}A. Moreo and D. J. Scalapino, 
                Phys.\ Rev.\ Lett.\ {\bf 66}, 949 (1991).
\bibitem{Mahan}G. D. Mahan, \em Many particle physics\em,
               2nd ed., Plenum Press, New York (1990).

\bibitem{Imptj} D.\ Poilblanc, D.\ J.\ Scalapino and
		W.\ Hanke, Phys.\ Rev.\ Lett.\ {\bf 72},
		884 (1994).
\bibitem{ImptjII} D.\ Poilblanc, D.\ J.\ Scalapino and
		W.\ Hanke, Phys.\ Rev.\ B {\bf 50},
		13020 (1994).
%
%
\bibitem{Hallberg} K.~A. Hallberg and C.~A. Balseiro,  Phys. Rev. B  {\bf 52},
                             374 (1995).

\end{thebibliography}
\end{document}